\documentclass[aps,prl,reprint,twocolumn,superscriptaddress,showpacs,amssymb]{revtex4-1}
\usepackage[english]{babel}
\usepackage{graphics}
\usepackage{graphicx}
\usepackage{epsfig}
\usepackage{amssymb}
\usepackage{dcolumn}% Align table columns on decimal point
\usepackage{bm}% bold math
\usepackage{color}
\usepackage{natbib}
\usepackage{lineno}
\usepackage{sidecap}
\usepackage{verbatim}  % Needed for the "comment" environment to make LaTeX comments
\usepackage{vector}  % Allows "\bvec{}" and "\buvec{}" for "blackboard" style bold vectors in maths
\usepackage{wrapfig}
\usepackage{enumerate}
\usepackage{sidecap}
\usepackage{amsmath}
\usepackage{hyperref}

\begin{document}

\title{Electronic Structure Studies of FeSi: A Chiral Topological System}

\author{Susmita Changdar}
\affiliation{Condensed Matter Physics and Material Science Department, S N Bose National Centre for Basic Sciences, 700106, India}
\author{S.\ Aswartham}
\affiliation{Leibniz Institute for Solid State Research, IFW Dresden, D-01171 Dresden, Germany}
\author{Anumita Bose}
\affiliation{Solid State and Structural Chemistry Unit, Indian Institute of Science, Bangalore 560012, India}
\author{Y. Kushnirenko}
\affiliation{Leibniz Institute for Solid State Research, IFW Dresden, D-01171 Dresden, Germany}
\author{G. Shipunov }
\affiliation{Leibniz Institute for Solid State Research, IFW Dresden, D-01171 Dresden, Germany}
\author{N. C. Plumb}
\affiliation{Paul Scherrer Institute, Swiss Light Source, CH-5232 Villigen PSI, Switzerland.}
\author{M. Shi}
\affiliation{Paul Scherrer Institute, Swiss Light Source, CH-5232 Villigen PSI, Switzerland.}
\author{Awadhesh Narayan}
\affiliation{Solid State and Structural Chemistry Unit, Indian Institute of Science, Bangalore 560012, India}
\author{B. B\"uchner}
\affiliation{Leibniz Institute for Solid State Research, IFW Dresden, D-01171 Dresden, Germany}
\author{S.\ Thirupathaiah}
\email{setti@bose.res.in}
\affiliation{Condensed Matter Physics and Material Science Department, S N Bose National Centre for Basic Sciences, 700106, India}
\affiliation{Leibniz Institute for Solid State Research, IFW Dresden, D-01171 Dresden, Germany}
\date{\today}

\begin{abstract}
Most recent observation of topological Fermi arcs on the surface of manyfold degenerate B20 systems, CoSi and RhSi, have attracted enormous research interests. Although an another isostructural system, FeSi, has been predicted to show bulk chiral fermions, it is yet to be clear theoretically and as well experimentally that whether FeSi possesses the topological surface Fermi arcs associated with the exotic chiral fermions in vicinity of the Fermi level. In this contribution, using angle-resolved photoemission spectroscopy (ARPES) and density functional theory (DFT), we present the low-energy electronic structure of FeSi. We further report the surface state calculations to provide insights into the surface band structure of FeSi near the Fermi level. Unlike in CoSi or RhSi, FeSi has no topological Fermi arcs near the Fermi level as confirmed both from ARPES and surface state calculations. Further, the ARPES data show spin-orbit coupling (SOC) band splitting of 40 meV, which is in good agreement with bulk band structure calculations.  We noticed an anomalous temperature dependent resistivity in FeSi which can be understood through the electron-phonon interactions as we find a Debye energy of 80 meV from the ARPES data.
\end{abstract}
\maketitle

$Introduction:-$ Since the discovery of linear dispersive Dirac fermions in graphene~\cite{Novoselov2005, CastroNeto2009}, the condensed matter has become fertile grounds for the investigation of various exotic topological fermions. Especially, the experimental realization of three-dimensional topological insulators~\cite{Hasan2010} further boosted the field to new heights, from basic sciences~\cite{Zhang2011b, Virot2011,Yan2012, Young2012, Wang2012a, Wang2013a,  Borisenko2014, Neupane2014a, Liu2014a, Huang2015,Zhang2015a,Ghimire2015,Shekhar2015, Xu2016, Tamai2016, Deng2016, Huang2016, Liang2016,Wang2016a, Jiang2017, Thirupathaiah2017, Thirupathaiah2017a, Thirupathaiah2018a} to more complex technological designs for the futuristic topological quantum computations (TQC)~\cite{Nayak2008, Lutchyn2010, Mourik2012, Stern2013, Stanescu2016}. At present, the topological quantum materials are classified by the Weyl fermions~\cite{Huang2015,Zhang2015a,Ghimire2015, Shekhar2015, Xu2016, Tamai2016, Deng2016, Huang2016, Liang2016,Wang2016a, Jiang2017, Haubold2017, Thirupathaiah2017, Thirupathaiah2017a}, the Dirac fermions~\cite{Young2012, Wang2012a, Wang2013a,  Borisenko2014, Neupane2014a, Liu2014a, Thirupathaiah2018a}, and the Majorana fermions~\cite{Sato2009a, Stanescu2016}. In general, at the band crossing point (BCP), the Weyl fermions have twofold degeneracy and the Dirac fermions have fourfold degeneracy. Recently, a new type of quantum materials have emerged with manyfold degenerate fermions at the band crossing point~\cite{Bradlyn2016, Tang2017, Chang2017, Pshenay-Severin2019, Rao2019, Takane2019}. These manyfold degenerate fermions are manifestations of the certain space-group symmetries in presence of the time-reversal invariance~\cite{Bradlyn2016}. Among them, the topological chiral systems like the transition-metal mono-silicides MSi (M = Co, Mn, Fe, Rh) have been at the recent intense research focus as under certain conditions, these systems are predicted to show spin-1/2 Weyl fermions with a topological charge of $\pm$ 1~\cite{Bradlyn2016, Chang2017, Tang2017, Shekhar2018, Takane2019, Sanchez2019, Yang2019, Schroeter2019, Rao2019}, spin-1 excitations with a topological charge of $\pm$ 2~\cite{Fang2012}, and spin-3/2 Rarita-Schwinger-Weyl (RSW) fermions with topological charges of $\pm4$~\cite{Rarita1941}. Moreover, the surface Fermi arcs connecting the manyfold degenerate BCPs are much longer in these systems compared to the other known Weyl and Dirac semimetals~\cite{Lv2015, Sun2015b, Xu2015, Deng2016}.

Earlier the transition metal monosilicides  were extensively studied for their low-energy electronic correlations~\cite{Fu1994, Park1995, Arko1997, Chernikov1997, Ishizaka2005,  Klein2008, Arita2008, Zhao2009, Petrova2010, Sales2011, Dutta2018}. Specifically, FeSi shows peculiar temperature dependent electronic and magnetic properties. It is an antiferromagnetic metal above 500 K, while a nonmagnetic narrow band gap insulator at low temperatures~\cite{Schlesinger1993, Jaccarino1967, Schlesinger1997, Takahashi1998, Fang2018}. Further, FeSi behaves as a semiconductor with an indirect band gap of 50 meV within the temperature range of 100-200 K~\cite{Schlesinger1993, Petrova2010},  while is a bad metal outside of this temperature range. Different mechanisms were proposed to explain this strange electronic and magnetic behaviour, (a) electron-phonon interactions~\cite{Delaire2011, Sales2011}, (b) spin fluctuations~\cite{Evangelou1983, Takahashi1997}, and c) charge excitations~\cite{Tomczak2012}. Apart from these interesting physical properties, FeSi is further predicted to show the above-mentioned manyfold degenerate chiral fermions at the high symmetry points with a nonzero Berry phase~\cite{Kuebler2013}. Further, a recent transport study on FeSi show anomalous temperature dependent resistivity which they attribute it to the plausible topological surface states~\cite{Fang2018}.

Motivated by the presence of surface Fermi arcs in RhSi and CoSi, we studied the low-energy electronic structure of isostructural FeSi using angle-resolved photoemission spectroscopy and density functional theory to show that despite FeSi is being chiral topological system, associated surface Fermi arcs connecting the manyfold degenerate bulk BCPs are absent near the Fermi level. These observations are further confirmed by our surface state calculations.  The ARPES data clearly show a spin-orbit coupling  band splitting of 40 meV, consistent with the theoretical calculations which predict a SOC split of 39.5 meV. We further noticed anomalous temperature dependent resistivity in FeSi, that means, FeSi is a semiconductor respecting the activation energy formula only within the temperature range of 75-143 K and eventually becoming a bad metal as moving away from the this temperature range. The spectral function analysis of ARPES data suggest an electron-phonon interaction at a Debye energy of 80 meV, while the spectral widths near the Fermi level changes merely by the thermal excitations within the range of 15-80 K.

%Since these exotic fermions are well above the Fermi level and due to the experimental limitations the band crossing points are not detected so far using ARPES.  However, one would detect the associated surface Fermi arcs is they exist in this system.
\begin{figure}[t]
\includegraphics[width=0.95\linewidth]{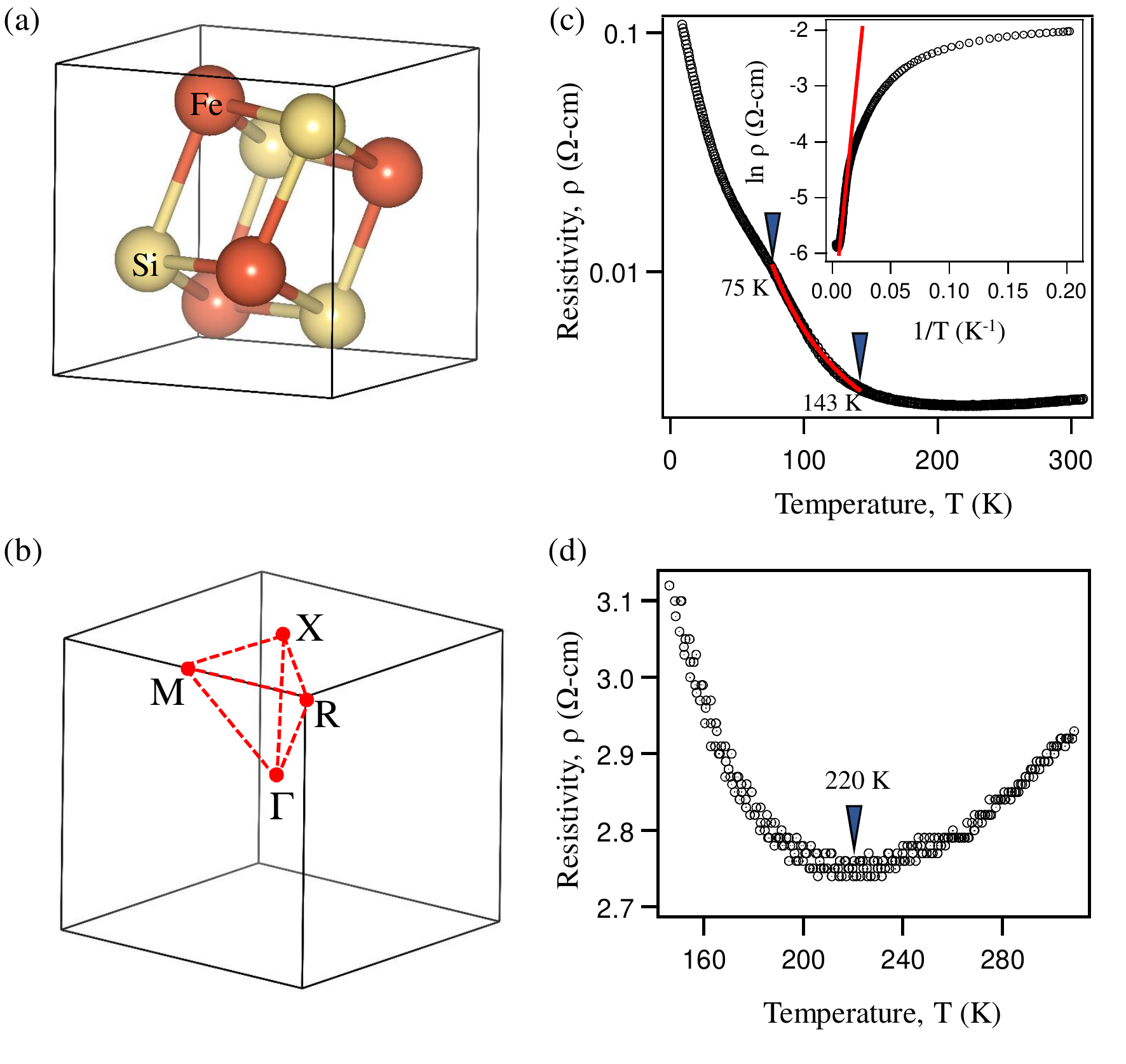}
\caption{(a) and (b) are the cubic crystal structure and corresponding Brillouin zone of FeSi, respectively. (c) Temperature dependent resistivity. Red curve in (c) is the activation formula fitting. Inset in (c) is the $\ln(\rho)$ $vs.$ 1/T. Red line in inset is the linear fitting.  (d) Zoomed in resistivity data at high temperature,  showing semiconductor to metal cross-over at T $\approx$ 220 K.}
\label{1}
\end{figure}

$Experimental ~details:-$ Single crystals were grown using floating zone method~\cite{Mason1979}. As grown single crystals were characterized using X-ray diffractometer (XRD) and energy dispersive X-ray analysis (EDAX).  These characterization techniques confirm the stoichiometric composition of FeSi and the space group of P2$_1$3 (198)~\cite{Damascelli1997, Petrova2010}.

Resistivity measurements were carried out on a closed cycle refrigerator (CCR) based cryostat of CRYOMECH.  Four copper (Cu) leads were connected to the sample by vacuum compatible silver epoxy (Epo-Tek H27D) in  Van der Pauw method. The sample temperature was varied between 4 and 320 K during the measurements.

ARPES measurements were carried out at 1$^3$-ARPES end station equipped with VG-Scienta R4000 electron analyzer in BESSY II (Helmholtz zentrum Berlin) synchrotron radiation center~\cite{Borisenko2012a, Borisenko2012b}. The angular resolution was set at $0.2^\circ$ for R4000.  Photon energies for the measurements were varied between 30 and 110 eV. The energy resolution was set between 10 and 15 meV depending on the excitation energy. Data were recorded at a chamber vacuum of the order of 1 $\times$ 10$^{-10}$ mbar and the sample  temperature was kept at 1 K during the measurements.  We employed various photon polarizations in order to extract the electronic structure comprehensively. Another set of ARPES measurements were performed in Swiss Light Source (SLS) at the SIS  beamline using a VG-Scienta R4000 electron analyzer.  Photon energy was varied between between 20 and 120 eV. Overall energy resolution was set between 15 and 25~meV depending on the photon energy.  Samples were cleaved $\textit{in situ}$ at a sample temperature of 15 K and the chamber vacuum was better than $5\times10^{-11}$ mbar during the measurements. At SIS  beamline, the data were recorded by varying the sample temperature between 15 and 80 K.

$Band ~structure ~calculations:-$ Band structure calculations were performed on the noncentrosymmetric cubic B20 crystal structure of FeSi~\cite{Brown1966},  having the lattice parameters of $a$ = $b$ = $c$ = 4.484 \AA, using density functional theory (DFT) within the generalized gradient approximation (GGA) of Perdew, Burke and Ernzerhof (PBE) exchange and correlation potential~\cite{Perdew1996} as implemented in the Quantum Espresso simulation package~\cite{QE-2009}. Ultra-soft non-relativistic and fully relativistic pseudopotentials were used to perform the calculations without spin-orbit coupling (SOC) and with SOC, respectively. The electronic wavefunction is expanded using plane waves up to a cutoff energy of 40 Ry (545 eV). Brillouin zone sampling is done over a 20$\times$20$\times$20 Monkhorst-Pack $k$-grid. The internal coordinates of the system are relaxed before producing the band structure.

For the surface state calculations, the tight-binding model was derived by computing the maximally-localized Wannier functions, choosing Fe $3d$ and Si $3p$ orbitals as the basis using the Wannier90 code~\cite{Mostofi2008}. We then employed WannierTools package~\cite{Wu2018} for analysis of surface and topological properties. Spin-orbit coupling was included for the surface calculations.

$Results ~and ~discussions:-$ Resistivity of FeSi as a function of temperature is shown in Figure~\ref{1}.  As can be seen from Figs.~\ref{1}(c) and \ref{1}(d),  the resistivity of FeSi decreases with increasing temperature up to 220 K, like a semiconductor. However, from a close observation we realise that FeSi is semiconductor only within the temperature range of 75-143 K as it can be properly fitted by the activation formula, $\rho(T)= \rho_{0}~e^{(\frac{E_{g}}{2k_{B}T})}$, where $E_{g}$ is the band gap. By fitting the resistivity data, as shown by the red line in the inset of Fig. \ref{1}(c), we estimate a semiconducting band gap of $E_g$=35 meV within this temperature range. The derived gap is in good agrement with previous report~\cite{Fu1994}. Further, we noticed a kink in the resistivity curve at around T = 75 K, below which $d\rho/dT$ decreases with the temperature. Similarly, we find that $d\rho/dT$ decreases with increasing temperature between 143 K and 220 K. And beyond 220 K, the resistivity increases with temperature. This peculiar resistivity character below 75 K and above 143 K can be attributed to bad metalicity of FeSi~\cite{Petrova2010, Fang2018}. Thus, our resistivity measurements suggest that FeSi is a semiconductor following the activation formula within the temperature range of 75-143 K and gradually becomes a bad metal as we go away from this temperature range. These results are qualitatively in agreement with the existing reports,  although the semiconducting temperature range is found to be different from different studies~\cite{Samuely1996, Buschinger1997, Paschen1997, Faeth1998, Sales2011, Fang2018}.

\begin{figure}[t]
\includegraphics[width=\linewidth]{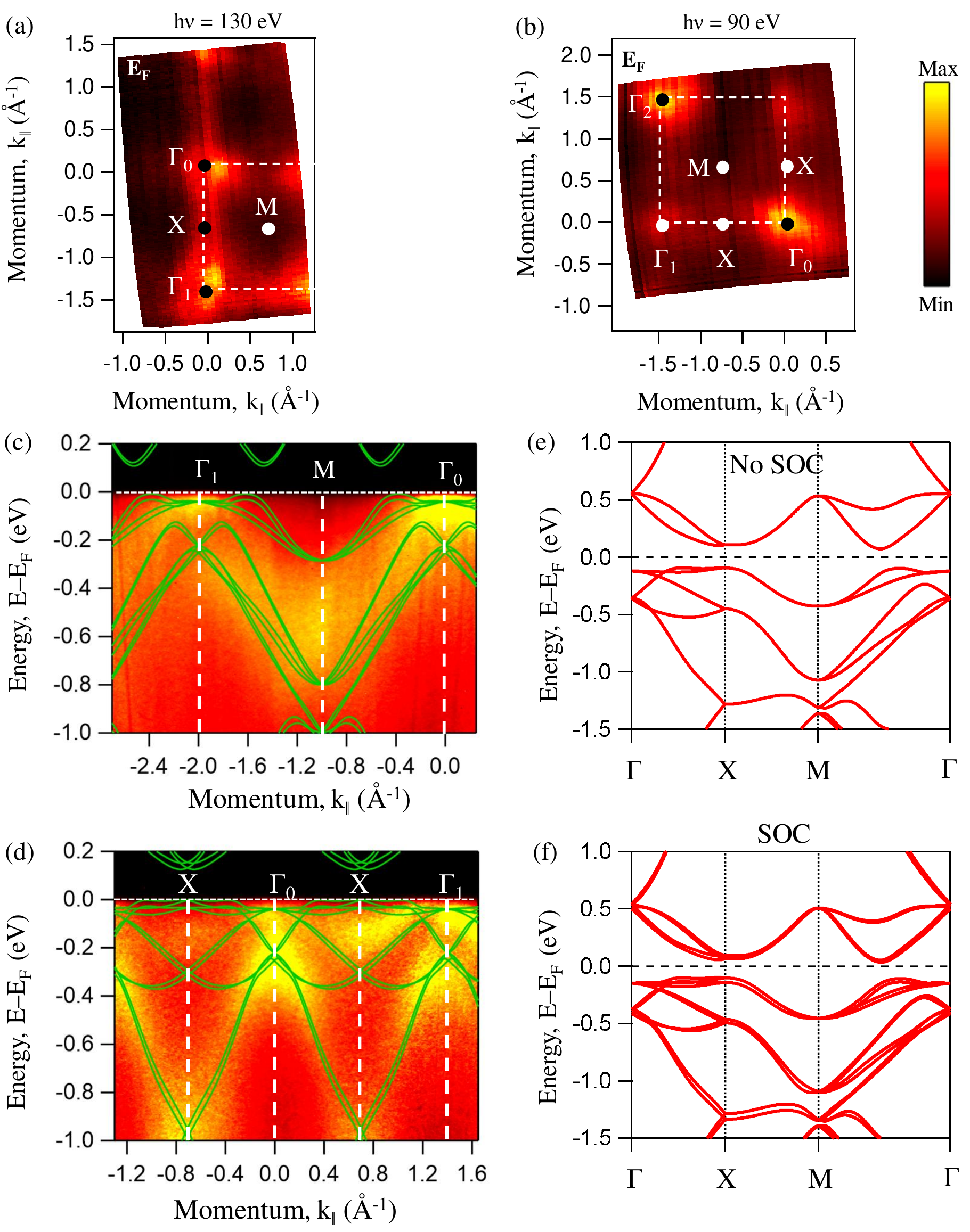}% Here is how to import EPS art
\caption{In-plane electronic band structure of FeSi. (a) and (b) are Fermi surface maps in $k_x-k_y$ plane measured using photon energy h$\nu$=130 eV and 90 eV, respectively. (c) Energy distribution map along the $\Gamma-M$ high symmetry line overlapped with DFT bulk band structure calculations including SOC.  (d) Energy distribution map along the $\Gamma-X$ high symmetry line overlapped with DFT band structure calculations including SOC. (e) DFT calculated band structure without SOC. (f) DFT calculated band structure with SOC.}
\label{2}
\end{figure}

Next, ARPES data of FeSi is shown in Figure~\ref{2} recorded at a sample temperature of 1 K. Fermi surface maps in the $k_x-k_y$ plane are shown in Figs.~\ref{2} (a) and ~\ref{2}(b) measured using $p$-polarized light with photon energies of 130 eV and 90 eV, respectively. Consistent with the crystal structure, the in-plane Fermi surface (FS) maps show the square symmetry. From the FS maps, we identify a blob-like spectral intensity  with fourfold symmetry at the $\Gamma$ point. On the other hand, we did not observe any clear spectral intensity either at $X$ or $M$ point when measured using $p$-polarized light. To further elucidate the nature of band dispersions, we show energy distribution maps (EDMs) along the high symmetry lines $\Gamma-M$ and $\Gamma-X$ as shown in  Figs.~\ref{2} (c) and ~\ref{2}(d), respectively,  measured using $p$-polarized light. DFT bulk band structure including spin-orbit coupling along the respective high symmetry lines is overlapped on to the EDMs. As can be seen from Figs.~\ref{2}(c) and ~\ref{2}(d), there is a good agreement between ARPES data and DFT calculations. Note here that the Fermi level of DFT calculations is shifted approximately 100 meV towards the higher binding energy to match with the experimental Fermi level. Band structure from the DFT calculations without SOC and with SOC in the $k$ path $\Gamma X M \Gamma$ are shown in Figs.~\ref{2}(e) and ~\ref{2}(f), respectively. Further,  using $s$-polarized light we could detect flat bands along the  $\Gamma-X$ high symmetry line (see Fig.~1 in supplementary information) which is in agreement with the DFT calculations. Thus, there is a finite spectral intensity at $X$ point that is clearly visible with the $s$-polarized light. This suggests that the band structure of FeSi near the Fermi level is composed by the multiple orbital characters. More details on the polarization dependent matrix elements can be found at Ref.~\onlinecite{Thirupathaiah2010}. Importantly, in Fig.~\ref{2}, we did not observe any spectral intensity related to the surface Fermi arcs.  Overall, the ARPES data shown in Fig.~\ref{2} supports the bad metallic picture of FeSi at low temperatures as observed from our resistivity measurements. Worth to mention here that, in Fig.~\ref{2}, for an easy representation,  we did not take into account the $k_z$ effects while assigning the high symmetry points on the Fermi surface maps and EDMs.

\begin{figure}[t]
    \centering
    \includegraphics[width=\linewidth]{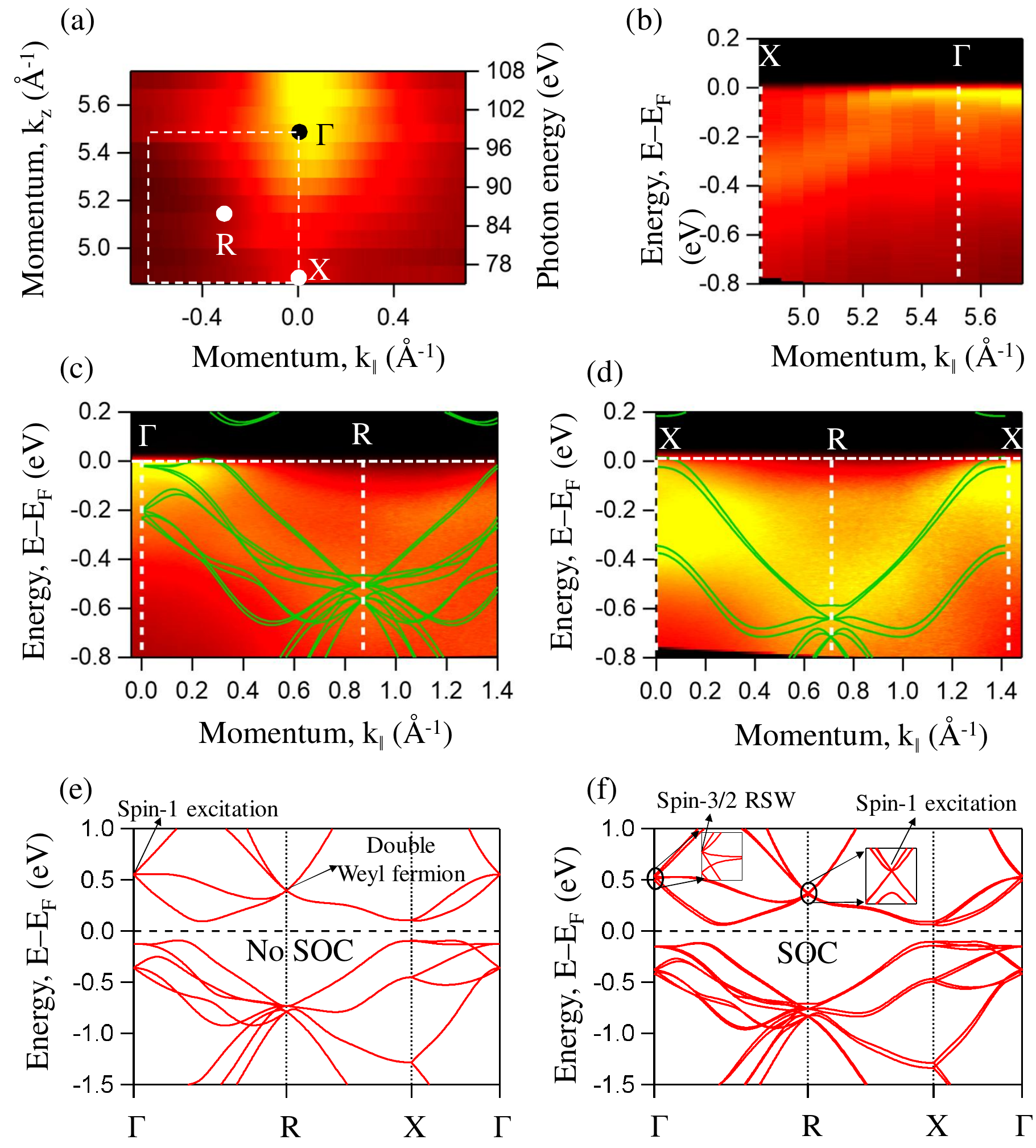}
    \caption{Out-of-plane electronic band structure of FeSi. (a) Fermi surface map in the $k_x-k_z$ plane. (b) Energy distribution map along $\Gamma$-X (001) high symmetry line. (c) Energy distribution map along the $\Gamma$-R high symmetry line overlapped with DFT band structure calculated including SOC. (d) Energy distribution map along the $X-R$ high symmetry line overlapped with DFT band struture calculated including SOC. (e) DFT band structure in the $\Gamma R X \Gamma$ without SOC and (f) with SOC.}
    \label{3}
\end{figure}

Photon energy dependent ARPES data are shown in Figure~\ref{3}. Fig.~\ref{3} (a) shows $k_y-k_z$ Fermi surface map measured with the photon energies ranging from 75 eV to 108 eV with a step of 3 eV using $p$-polarized light. The high symmetry points $\Gamma$, $X$ and $R$ are denoted on the Fermi surface map following the equation, $k_z~=~\sqrt{\frac{2m}{\hbar^2} ({V_0 + E_k\cos^2\theta})}$ with an inner potential of 16 eV. From the $k_z$ Fermi surface map we realize that the photon energy of 100$\pm$3 eV detects the bands from the $\Gamma$ point and photon energy of 75$\pm$3 eV detects the bands from the $X$ point. Similarly, the high symmetry point $R$ is accessible with a photon energy of 86$\pm$3 eV when the sample surface is normal to the $c$ axis. Energy distribution maps along $\Gamma-X$, $\Gamma-R$ and $X-R$ are shown in Figs.~\ref{3}(b), (c), and (d), respectively. The band structure extracted along the in-plane $\Gamma-X$ ([100]) as shown in Fig.~\ref{3}(b), is in good agreement with the band structure extracted along the out-of-plane $\Gamma-X$ ([001]) as shown in Fig.~\ref{2}(d). The band structure derived from DFT calculations with SOC along $\Gamma-R$ and $X-R$ is overlapped on the experimental band structure as shown in Figs.~\ref{3}(c) and (d), and there is a good agreement between DFT calculations and ARPES data. The calculated bulk band structure without SOC and with SOC in the $k$-path $\Gamma R X \Gamma$ are shown in Figs.~\ref{3}(e) and (f), respectively. As predicted from the DFT calculations without SOC, in FeSi the triple-point spin-1 excitations with topological charge of $\pm$2 are at the $\Gamma$ point and double Weyl fermions with topological charge of $\pm$2 are at the $R$ point. On the other hand, DFT with SOC,  the triple-point spin-1 excitations are predicted at the $R$ point while the spin-3/2 Rarita-Schwinger-Weyl fermions are predicted at the $\Gamma$ point~\cite{Tang2017}.

\begin{figure}[b]
    \centering
    \includegraphics[width=\linewidth]{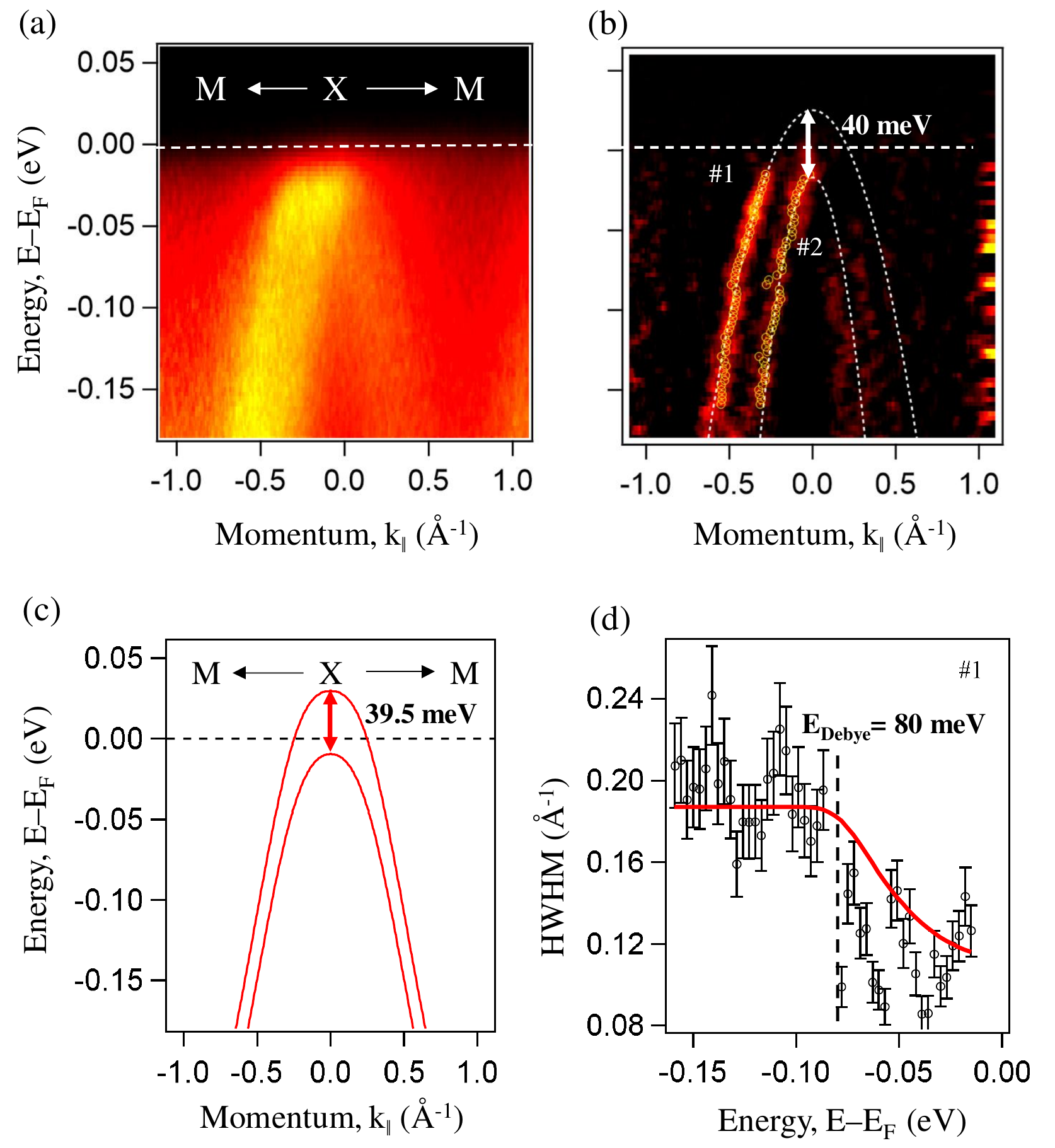}
    \caption{Temperature dependent ARPES data. (a) Energy distribution map along the $X-M$ high symmetry line measured using 90 eV photon energies at a sample temperature of 15 K. (b) Second derivative of (a). White dashed curves in (b) are the parabolic fits to the experimental bands.  (c) Zoomed in DFT band structure along $X-M$ orientation. (d) Half width halm maximum (HWHM) as a function of binding energy extracted from the EDM shown in (a) by fitting the momentum dispersive curves (MDCs) with the Lorentzian function. In (d),  the red curve is a fit of the self-energy function.}
    \label{4}
\end{figure}

Figure~\ref{4}(a) depicts EDM along the $X-M$ orientation measured at a sample temperature of 15 K.  From the second derivative of Fig.~\ref{4}(a) as shown in  Fig.~\ref{4}(b) we identify two band dispersions, \#1 and \#2. Here, the band \#1 is crossing the Fermi level with a momentum vector of 0.22 $\AA^{-1}$ at the $X$ point whereas the band \#2 does not cross the Fermi level. Further, we estimate an energy difference between the top of \#1 and \#2 is about 40 meV, which is in good agreement with the DFT calculations with SOC which predicts it to be 39.5 meV as shown in Fig.~\ref{4}(c). In fact, without SOC there exists only one band dispersion along $X-M$ orientation at this energy position [see Fig.~\ref{2}(e)]. Thus, the experimental band structure can be properly reproduced using DFT calculations only with the SOC inclusion. To further elucidate temperature effects on the electronic band structure of FeSi, we measured EDMs along $X-M$ orientation with temperature ranging between 15 K and 80 K (see Fig.~2 in the supplementary information). From the temperature dependent EDMs as shown in Fig.~2 of the supplementary information, it is evident that the band structure near the Fermi level hardly changes with the temperature at least within the range of 15-80 K.  We further estimated half-width-half-maximum (HWHM) from MDCs which is directly related to the imaginary part of the self-energy ($\Im\Sigma(E)$) for the band \#1 as shown in Fig.~\ref{4}(d). By fitting HWHM using the self-energy function~\cite{Valla1999}, we find electron-phonon coupling at a Debye energy of 80 meV. This estimate of  Debye energy is in good agreement with an earlier ARPES report which suggested a Debye energy of 90 meV~\cite{Klein2008}. Thus, the anomalous resistivity observed in FeSi (see Fig.~\ref{1}) may not be of the electronic structure origin. But, based on the spectral functional analysis,  we suggest that the electron-phonon coupling is playing a crucial role for the observed anomalous resistivity also as suggested by the previous reports~\cite{Racu2007, Menzel2009, Delaire2011, Sales2011}.

%With the help of temperature dependent EDMs we estimated spectral widths of bands \#1 as a function of temperature at 50 meV below the Fermi level as shown in Fig.~\ref{4}(d). Fig.~\ref{4}(e) shows temperature dependent energy distribution curves taken from an integration over the momentum window of 0.5 $\AA^{-1}$ about $k=0$.

\begin{figure}[bp]
    \centering
    \includegraphics[width=\linewidth]{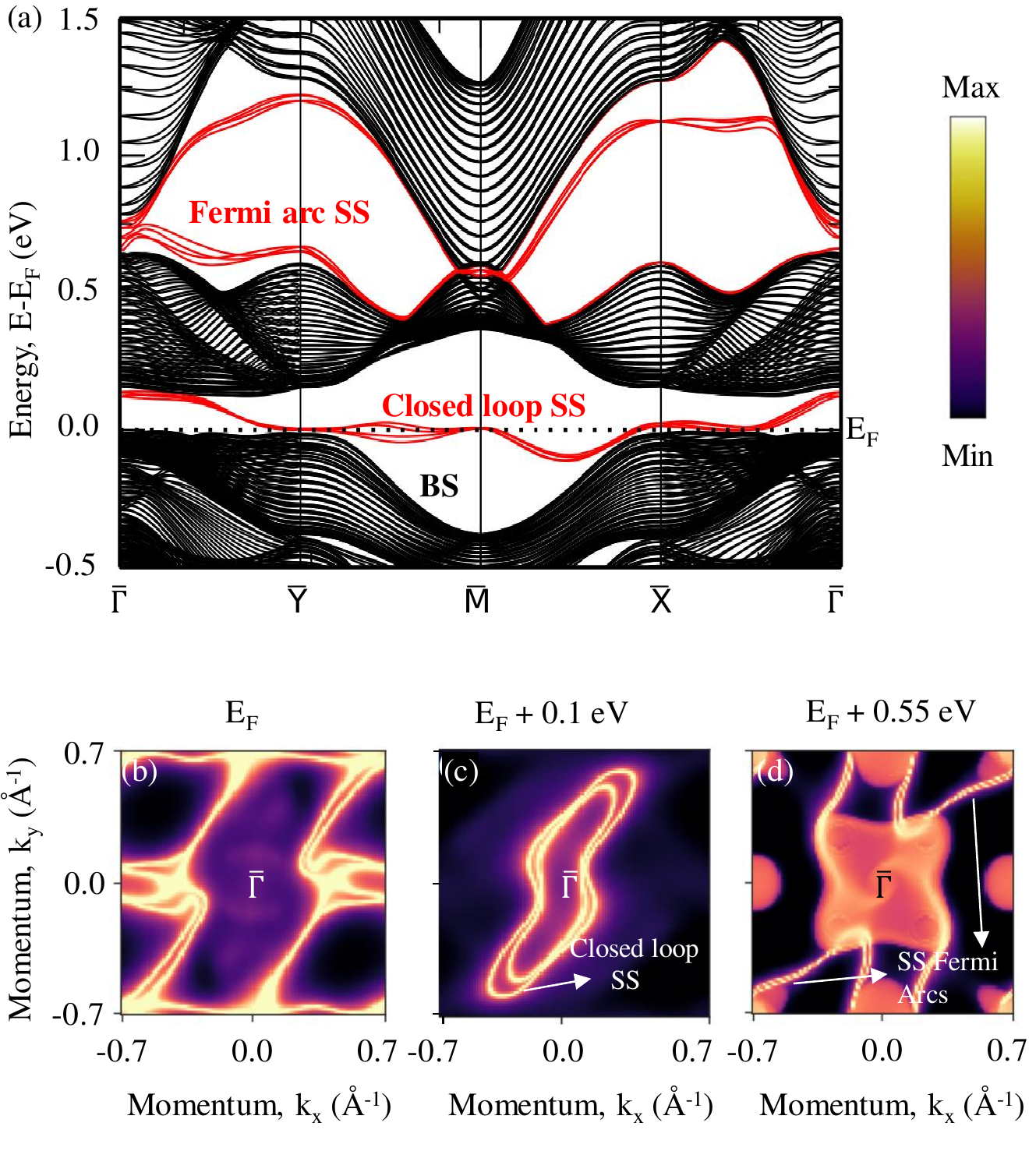}
    \caption{Band structure calculated for a (001) oriented FeSi slab of thickness 88.96$\AA$. (a) Energy-momentum plot showing surface states (red colored) overlapped with the bulk band structure. Surface state Fermi map taken at E$_F$ (b), E$_F$+0.1eV (c), and E$_F$+0.55eV (d).}
    \label{5}
\end{figure}

Overall, the experimental band structure is quantitatively in agreement with the DFT calculations. Specially, the experimental observation of SOC band splitting has been explained very well from DFT calculations with spin-orbit interactions. Till date,  a very few ARPES data with low energy and momentum resolution are available in the literature on FeSi, so it is difficult to compare quantitatively. However, qualitatively, our ARPES data is consistent with some of the earlier ARPES reports~\cite{Arita2008, Klein2008}. Next, coming to the main point of this manuscript, recent ARPES reports on CoSi and RhSi showed topological surface Fermi arcs spanned over a large area of 2D Brillouin zone~\cite{Chang2017, Tang2017, Shekhar2018, Takane2019, Yang2019,Sanchez2019, Schroeter2019, Rao2019, Takane2019}. Moreover, they could record manyfold bulk Weyl fermions at $\Gamma$ and $R$ high symmetry points. In contrast, we could not observe any such surface Fermi arcs from our ARPES measurements performed on the isostructural FeSi. As predicted by the DFT calculations [see Fig.~\ref{3}(f)], in FeSi,  the manyfold spin-3/2 RSW fermions are at 0.54 eV and the triple-point spin-1 excitations are at 0.37 eV above the Fermi level. So, it would not be possible to measure them using ARPES technique. Nevertheless, based on the present understanding,  irrespective of the energy position of the manyfold band crossing points (BCPs) one would expect the associated Fermi surface arcs on the surface Brillouin zone~\cite{Sun2015a, Bradlyn2016, Wang2016b}.

Therefore, to gain more insight into the surface band structure of FeSi, we carried out density functional theory calculations as shown in Fig.~\ref{5}. We constructed a wannier function based model to compute the band structure of FeSi slab oriented along the [001] direction. The band structure, for a slab of thickness 88.96 $\AA$, along the high symmetry directions is shown in Fig.~\ref{5}(a). Most noticeably, we find only a set of trivial surface bands within the bulk band gap, and the absence of any topological protected Fermi arcs close to the Fermi level. The topological Fermi arcs, associated with manyfold fermions similar CoSi and RhSi, occur at substantially higher energies (0.55 eV above the Fermi level). Furthermore, we also considered the semi-infinite geometry, employing a Green's function method to calculate the surface states, as a function of the in-plane momenta, at diﬀerent energies. These are presented in Fig.~\ref{5}(b)-(d). In stark contrast to the case of CoSi, we find that these surface states near the Fermi level close-in on themselves as shown in Figs.~\ref{5}(b) and ~\ref{5}(c), clearly indicating the triviality of these states. The reciprocal space extent of these surface states diminishes as one moves away from the Fermi energy, with the closed loops shrinking in size. However, though the shape of the Fermi arcs is a bit different from CoSi and RhSi,  the non-trivial topological Fermi arcs can be noticed in FeSi at 0.55 eV above the Fermi level as shown in Fig.~\ref{5}(d). Thus the surface state calculations indicate that the topological Fermi arcs present in FeSi, however, they are not accessible by conventional ARPES technique. Further, these calculations predicted trivial surface states near the Fermi level which are not well resolved in our ARPES data due to either the surface state spectral intensity is very low compared to the bulk spectral intensity or the sample surface quality is not good enough to detect them.

%Therefore, presuming that the band crossing points away from the Fermi level in FeSi could not effect for the absence of surface Fermi arcs.   Further, our ARPES studies do not support the recent suggestion of surface states in FeSi made using the transport studies~\cite{Fang2018}.

$Conclusions:-$
We systematically studied the low-energy electronic structure of topological chiral fermionic system, FeSi, using angle-resolved photoemission spectroscopy and density functional theory to derive the following conclusions,

1. Observation of Fermi surface from the ARPES measurements suggest that FeSi is a metal at low temperature, in agreement with our resistivity measurements.

2. ARPES data show a spin-orbit band splitting of 40 meV that is nicely reproduced by the DFT calculations including SOC. Therefore,  SOC effects must be considered while discussing the physics of manyfold degenerate fermions in the transition metal monosilicides.

3. Anomalous temperature dependent resistivity of FeSi can be explained by the electron-phonon interactions.

4. Unlike in the case of CoSi or RhSi, FeSi do not show topological surface Fermi arcs near the Fermi level as surface state calculations predicted them well above the Fermi level. Therefore, we are unable to detect them using conventional ARPES technique.

$Acknowledgements:-$
S.C. acknowledges University Grants Commission (UGC), India for the PhD fellowship. A.B. thanks Indian Institute of Science for the PhD fellowship. A.N. acknowledges support from the start-up grant at the Indian Institute of Science. S.A. acknowledges financial support by DFG through grant AS 523/4-1. S.A. and B.B. thank DFG through grant AS 523/3-1.   S.T. acknowledges the financial support by DST, India through the INSPIRE-Faculty program (Grant No. IFA14 PH-86). S.T. acknowledges the financial support given by SNBNCBS through the Faculty Seed Grants program. S.T. Acknowledges the travel support given by DST, India (SR/NM/Z-07/2015) and Jawaharlal Nehru Centre for Advanced Scientific Research (JNCASR) for managing the project. The authors thank Alexander Fedorov and Emile Rienks for their technical support during the experiments performed at BESSYII (HZB).

\bibliography{FeSi}

\section{Supplementary information}

\begin{figure*}[htbp]
    \centering
    \includegraphics[width=\linewidth]{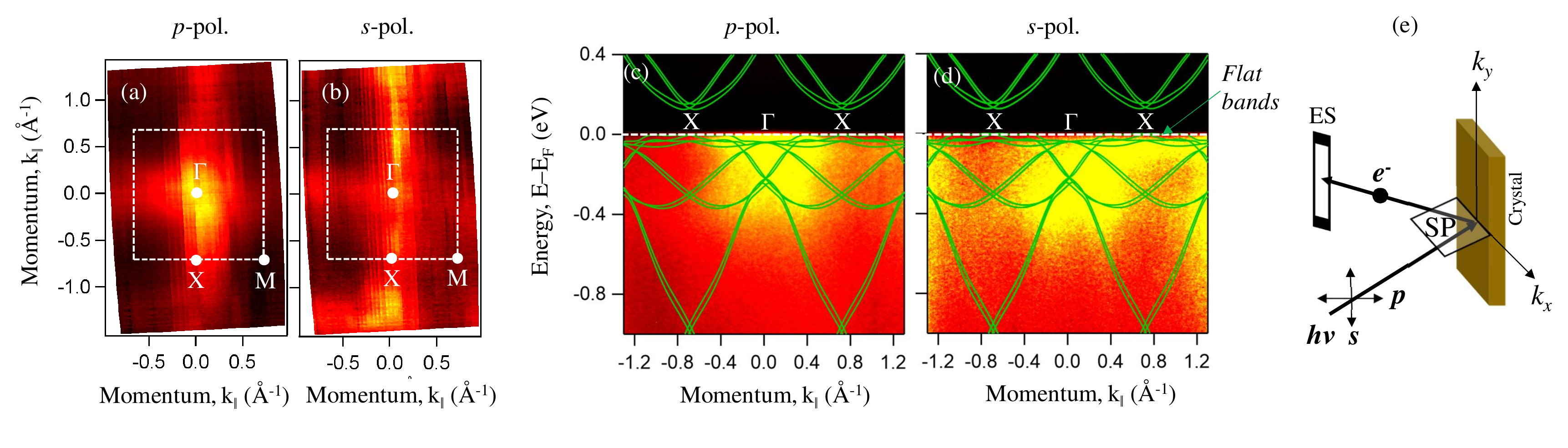}
    \caption{Polarization dependent Fermi surface maps measured with a photon energy of 100 eV. (a)  $p$-polarization. (b) $s$-polarization. (c) and (d) are the EDMs taken along the $\Gamma-X$ orientation using $p$- and $s$- polarizations, respectively. (e) ARPES measurement geometry in which the $s$ and $p$ polarizations are defined with respect to the scattering plane (SP).}
    \label{S1}
\end{figure*}

\begin{figure*}[htbp]
    \centering
    \includegraphics[width=\linewidth]{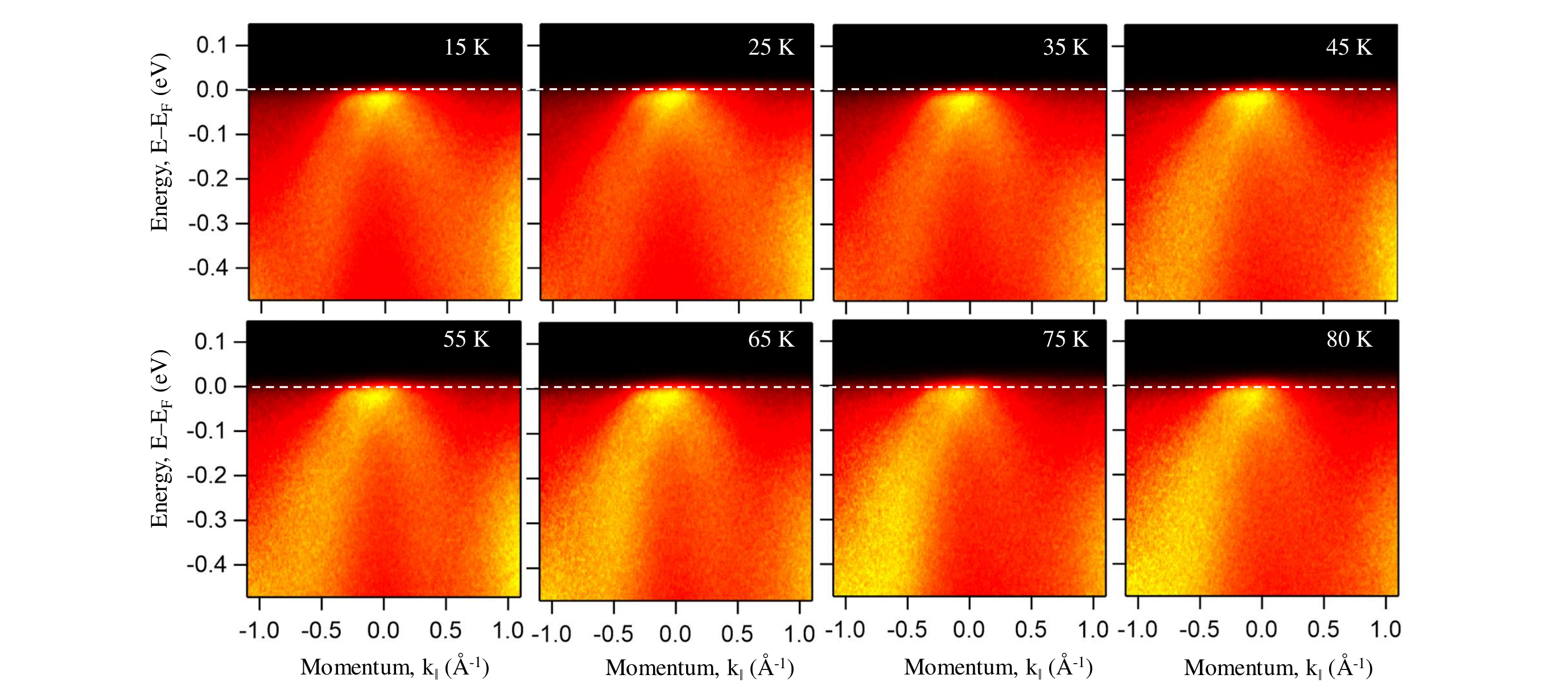}
    \caption{Temperature dependent ARPES data measured using $s$-polarized light with a photon energy of 90 eV.}
    \label{S2}
\end{figure*}

\end{document}